\documentclass[aps,prb,preprint]{revtex4-1}
\usepackage{epsfig,graphicx}
\usepackage{amsmath}
\usepackage{amsfonts}
\usepackage{amssymb}
\usepackage{color}
\usepackage{ulem}
\begin{document}

\title{Ab-initio perspective on the Mollwo--Ivey relation for F-centers in alkali halides}

\author{Paul Tiwald}
\affiliation{Institute for Theoretical Physics, Vienna University of Technology, Wiedner Hauptstra\ss e 8-10, A-1040 Vienna, Austria, EU}

\author{Ferenc Karsai}
\affiliation{Institute of Materials Chemistry, Vienna University of Technology, Getreidemarkt 9/165-TC, A-1060 Vienna, Austria, EU}

\author{Robert Laskowski}
\affiliation{Institute of High Performance Computing,
A*STAR 1 Fusionopolis Way, \# 16-16, Connexis, Singapore 138632}
\affiliation{Institute of Materials Chemistry, Vienna University of Technology, Getreidemarkt 9/165-TC, A-1060 Vienna, Austria, EU}

\author{Stefanie Gr\"afe}
\affiliation{Institute for Theoretical Physics, Vienna University of Technology, Wiedner Hauptstra\ss e 8-10, A-1040 Vienna, Austria, EU}
\affiliation{Institute of Physical Chemistry and Abbe Center of Photonics, Friedrich-Schiller-University Jena, Helmholtzweg 4, D-07743 Jena, Germany, EU}

\author{Peter Blaha}
\affiliation{Institute of Materials Chemistry, Vienna University of Technology, Getreidemarkt 9/165-TC, A-1060 Vienna, Austria, EU}

\author{Joachim Burgd\"orfer}
\affiliation{Institute for Theoretical Physics, Vienna University of Technology, Wiedner Hauptstra\ss e 8-10, A-1040 Vienna, Austria, EU}

\author{Ludger Wirtz}
\affiliation{Physics and Materials Science Research Unit, University of Luxembourg, 162a avenue de la Fa\"iencerie, L-1511 Luxembourg, Luxembourg, EU}
\affiliation{Institute for Electronics, Microelectronics, and Nanotechnology (IEMN), CNRS UMR 8520, Dept.\ ISEN, F-59652 Villeneuve d'Ascq Cedex, France, EU}

\begin{abstract}
We revisit the well-known Mollwo--Ivey relation that describes the ``universal'' dependence of the absorption energies of F-type color centers on the lattice constant $a$ of the alkali-halide crystals, $E_{\mbox{abs}}\propto a^{-n}.$ We perform both state-of-the-art ab-initio Quantum Chemistry and post-DFT calculations of F-center absorption spectra. By ``tuning'' independently the lattice constant and the atomic species we show that the scaling of the lattice constant alone (keeping the elements fixed) would yield $n=2$ in agreement with the ``particle-in-the-box'' model. Keeping the lattice constant fixed and changing the atomic species enables us to quantify the ion-size effects which are shown to be responsible for the exponent  $n \approx 1.8$.
\end{abstract}


\maketitle

\section{Introduction}
\label{sec:introduction}
Exploration of the physics of the color centers has a long history dating back to the late 1920s. Still, color centers constitute  an important topic of current research as they offer optically addressable quasi-atomic localized excitations in the condensed phase only weakly coupled to a decohering environment. Prominent examples include nitrogen-vacancy color centers as candidates for the realization of quantum bits\cite{NemTruDevSte14}. Color centers play a crucial role in various optical devices such as tunable solid-state lasers\cite{TerTsu96} (see also ref.\ \onlinecite{ChiBonGomMic14} and references therein). Furthermore, the coupling of the electronic and nuclear degrees of freedom in optical absorption and emission processes of color centers is of interest for resonance spectroscopies\cite{KoySue11} and coherent state manipulation. Meanwhile a considerable number of ab-initio studies of the electronic structure of color centers have become available\cite{CarSouIllSus06,MaRoh08,RinSchKioJan12,KarTiwLasTra14}.
\\
The prototypical color center, first investigated in the late 1920s, is the F center in alkali-halide crystals with rock-salt structure. It consists of a single electron captured in an anion vacancy. The ground state of the electron residing near the defect is localized in the vacancy with a nearly spherical, or s-type, wave function. Upon absorption of light the defect electron is excited in a dipole transition into a p-type state. In a recent work\cite{KarTiwLasTra14} we have analyzed in detail this quasi-atomic transition for lithium fluoride (LiF) employing complementary state-of-the-art ab-initio methods from quantum chemistry and solid-state theory.
\\
In the present work we extend this analysis to other alkali halides and address the physics underlying the well-known Mollwo--Ivey relation. Mollwo \cite{Mol31} pioneered the observation that the absorption energy $E_{abs}$ of F centers in alkali halides scales like $\sim 1/a^2$, where $a$ is the lattice constant (i.e., twice the anion-cation distance). Later, Ivey \cite{Ive47} found, based on a larger data set for different alkali halides, that $E_{abs}$ of many defects follow
\begin{align}
\label{eq:MI}
E_{abs} = C a^{-n},
\end{align}
where $C$ is a proportionality constant and $n$ is the so-called Mollwo--Ivey exponent. Fitting this expression to experimental data of F-center absorption energies in crystals with rock-salt structure \cite{DawPoo69,MalSmi92} yields $n=1.81 \pm 0.1$ with a universal prefactor $C=17.3 \pm 2.8$~eV (when $a$ is given in \AA ngstr\"om). These findings stimulated a large number of theoretical investigations over the years\cite{Fro33,Sto51,Sto52-1,Sto52-2,Woo65,Fow68} resulting in a wide variety of qualitative and semi-quantitative models aimed at explaining this remarkably simple scaling relation. While in detail substantially different, the common idea underlying these models is the notion of the ``particle in the box'': an electron is trapped at the site of a missing anion in order to render the crystal neutral. This electron cannot move through the crystal due to the large band gap of alkali halides. In the 1960s, more experimental data, in particular the observed pressure dependence in heavier alkali halides\cite{BucFit68}, appeared to deviate from Eq.\ \ref{eq:MI} and refinements were proposed to account for so-called ion-size effects\cite{BucFit68,BarStoGas68,MalSmi92,SmiMal98} that are not expected to follow any simple $a$-scaling.
\\
A systematic investigation of Eq.\ \ref{eq:MI} based on state-of-the-art ab-initio simulations appears to be still missing. We therefore present calculations of the defect wave functions and absorption energies of F centers for several alkali-halide crystals. Our calculations shed light on the exact nature of the electron confinement which is mainly due to the exchange interaction and the requirement of orthogonality of the defect wave function and the ion core levels. By independently varying the lattice constant and the ionic species, we can disentangle the effects of the scaling with lattice size from that of the influence of effective ionic radii.
\\
The paper is structured as follows. In sec.\ \ref{sec:methods} we briefly review the computational methods employed in the present study. In section \ref{sec:models} we give a brief overview over previously proposed models for explaining the Mollow--Ivey relation and scrutinize them in the light of the current results. In Sec.\ \ref{sec:scaling} we introduce the scaled alkali halide model which allows to disentangle lattice-size from ion-size effects. Ion-size effects are studied in section \ref{sec:ion-size-effects} and their influence on the absorption energies and the Mollwo--Ivey exponent are presented in section \ref{sec:energies}. Conclusions and an outlook are given in section \ref{sec:conclusions}. 

\section{Computational Methods}
 \label{sec:methods}
The numerical results presented in the following are based on two complementary state-of-the-art ab-initio approaches. On the one hand, we employ quantum-chemistry methods for localized molecule-like systems. On the other-hand, we use post-density functional theory (post-DFT) methods for extended condensed matter systems (for details see Ref.\ \onlinecite{KarTiwLasTra14}).
\\
Briefly, quantum-chemistry techniques are employed within the framework of the so-called embedded cluster approach\cite{deGSouBro98} (ECA). Due to the strong localization of the F-centers the defect and its environment can be represented by a finite-size active cluster. Unless otherwise stated we use an active cluster containing 18 anions and 38 cations and the defect, i.e.\ a missing anion in the very center. For this cluster size, calculations of the F-center absorption energies in alkali halides with large elements such as potassium chloride are still feasible, defect wave functions are well converged, and absorption energies are converged up to $\sim 0.2$~eV. The active cluster is embedded into several layers of \textit{ab initio} model potentials \cite{HuzSeiBarKlo87,SeiBar99} (AIMPs) and a large matrix of point charges of cubic shape arranged as proposed by Evjen\cite{Evj32} with fractional charges of +/- 0.5, 0.25, and 0.125 at faces, edges, and corners, respectively. All active atoms/ions, AIMPs, and point charges are placed at lattice positions of the pristine alkali-halide crystals, i.e., for the study of the universal $a$-scaling for alkali halides (Eq.\ \ref{eq:MI}, Sec.\ \ref{sec:scaling}) we deliberately omit an element-specific geometry optimization. We consider relaxation effects when we discuss detailed comparison with experimental data (sec.\ \ref{sec:energies}). We use Dunning's correlation-consistent polarized valence-only basis set\cite{Dun88} of triple zeta quality (cc-pVTZ) for the active clusters. Since no such basis set is available for potassium, we use for alkali halides involving K the small Atomic Natural Orbital (s-ano) basis set\cite{PieDumWidRoo95}, which is of comparable size. For those crystals for which both basis sets are available we find hardly any difference in the absorption energy for the s-ano and the cc-pVTZ basis set. One- and two-electron integrals are computed using the Cholesky decomposition\cite{AquDeVFerGhi10} as implemented in the \textsc{Molcas} program package.\\
F-center absorption energies are determined as differences between total energies of the ground and first optically allowed excited N-electron state of the active cluster. The total energies are calculated using the complete active space second-order perturbation theory (CASPT2) based on restricted open-shell Hartree--Fock (ROHF) wave functions. This method has been shown to account for correlation effects to a large degree \cite{KarTiwLasTra14} as dynamical correlations strongly dominate over static correlations for wide-band gap insulators\cite{GruMarKre10}. All occupied HF orbitals are doubly occupied and localized at the active anions, i.e.\ halides, and cations, i.e.\ alkali metals, except for the defect orbital which is well localized in the vacancy region and hosts the unpaired F-center electron. This orbital has s and p-type character\cite{KarTiwLasTra14} in the ground and excited state, respectively, and its size and shape are well converged for the cluster and basis set size used. Its representation by one-electron orbitals has been shown to be remarkably accurate by comparison with the natural-orbital representation of the reduced density matrix of the full N-electron wave function\cite{KarTiwLasTra14}.\\
The solid-state approach starts with DFT calculations for periodic super cells using the  WIEN2k code~\cite{WIEN2k_ref}, which is based on the full-potential (linearized) augmented plane-wave ((L)APW) + local orbitals (lo) method. In contrast to the quantum chemistry calculations we use a basis set consisting of a combination of plane waves and atomic-like basis sets localized in spheres around the ionic cores. We use in all calculations unit cells containing 31 atoms, a $k$-mesh size of $3\times 3\times 3$ and a RKMAX=7.0, where RKMAX is the product of the smallest atomic sphere radius (1.57, 2.04, 2.5~bohr for Li, Na, K and 2.02, 2.5, 2.6 for F, Cl, Br, respectively) times the plane wave basis set cutoff KMAX. The required unit cell sizes and the Brillouin zone sampling ($k$ mesh) are tested for convergence.\\
All calculations are carried out spin-polarized. The structure relaxations are performed using the PBE exchange-correlation functional~\cite{PBE_ref}. Since PBE strongly underestimates the band gap in most materials~\cite{BGPROB_1,BGPROB_2,BGPROB_3} (also known as the "band-gap problem"), we calculate the electronic structure with the TB-mBJ potential~\cite{TB-mBJ_1}, which is parameterized to give accurate band gaps in semiconductors and insulators~\cite{TB-mBJ_2,TB-mBJ_3}. The TB-mBJ potential method can be viewed as an alternative to the very time-consuming $GW$ approach~\cite{GW_ref} giving almost identical results for the $F$-center in LiF. Due to the strong localization of the $F$-center electron, excitonic binding energies of several eV are expected. In contrast to the quantum chemistry calculations these electron-hole correlation effects are not included in standard band-structure calculations. To account for them we solve the Bethe-Salpeter equation~\cite{BSE_1,BSE_2,BSE_3} (BSE) on top of the TB-mBJ calculations when calculating the absorption spectra.

\section{Models for F centers and the confining potential}
\label{sec:models}
\begin{figure}[htb]
\centering \includegraphics[angle=0,width=0.5\columnwidth]{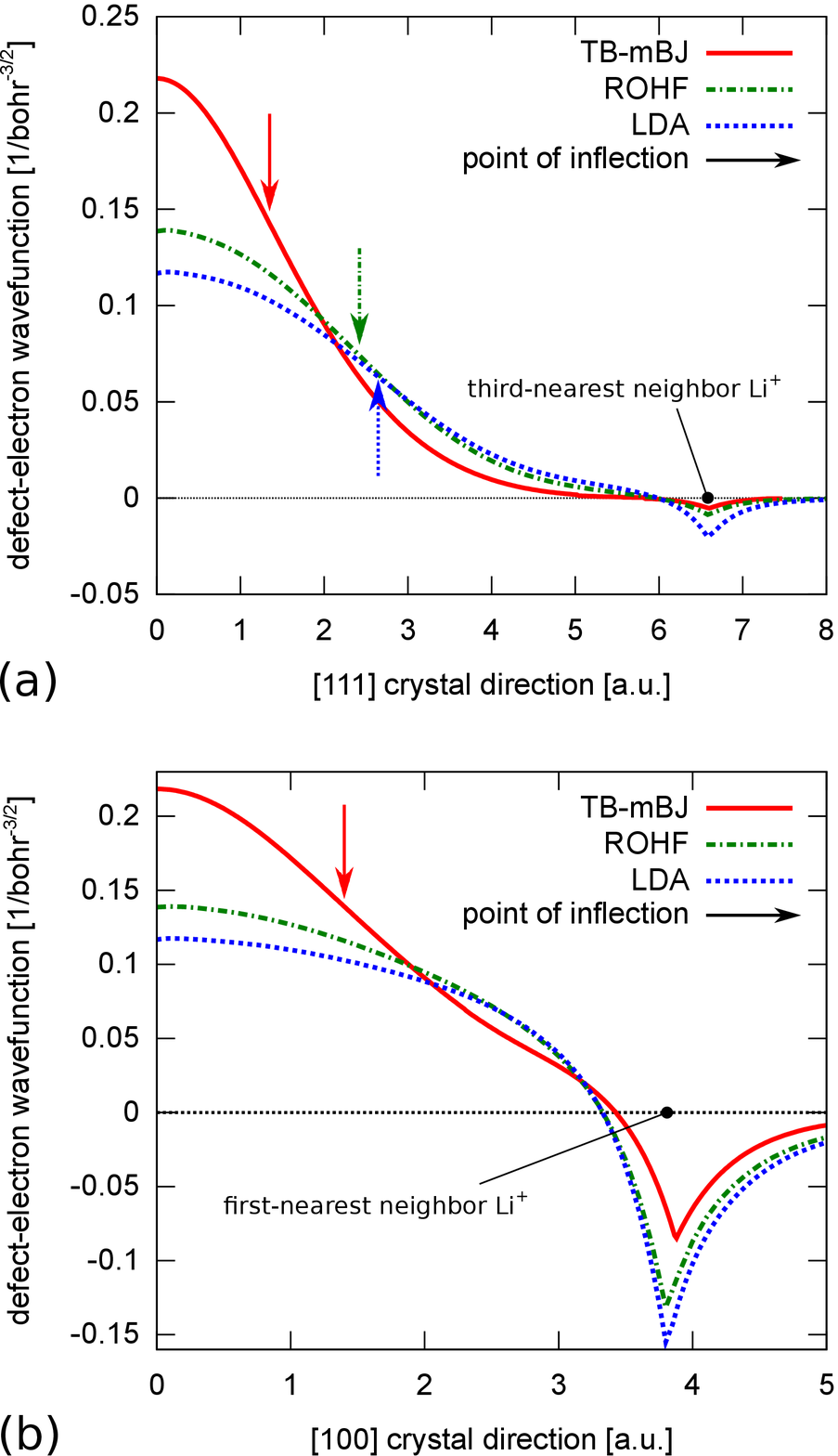}
\caption{(Color online) (a) Cut through the ground state TB-mBJ (solid red line), ROHF (dash-dotted green), and LDA (blue dotted) orbital of the F-center electron in LiF along the (a) [111] and (b) [100] crystal directions. Vertical arrows indicate point of inflections where the wave function enters a classically forbidden region and penetrates the effective potential barrier.}
\label{fig:pofI}
\end{figure}

\begin{figure}[htb]
\centering \includegraphics[angle=0,width=0.5\columnwidth]{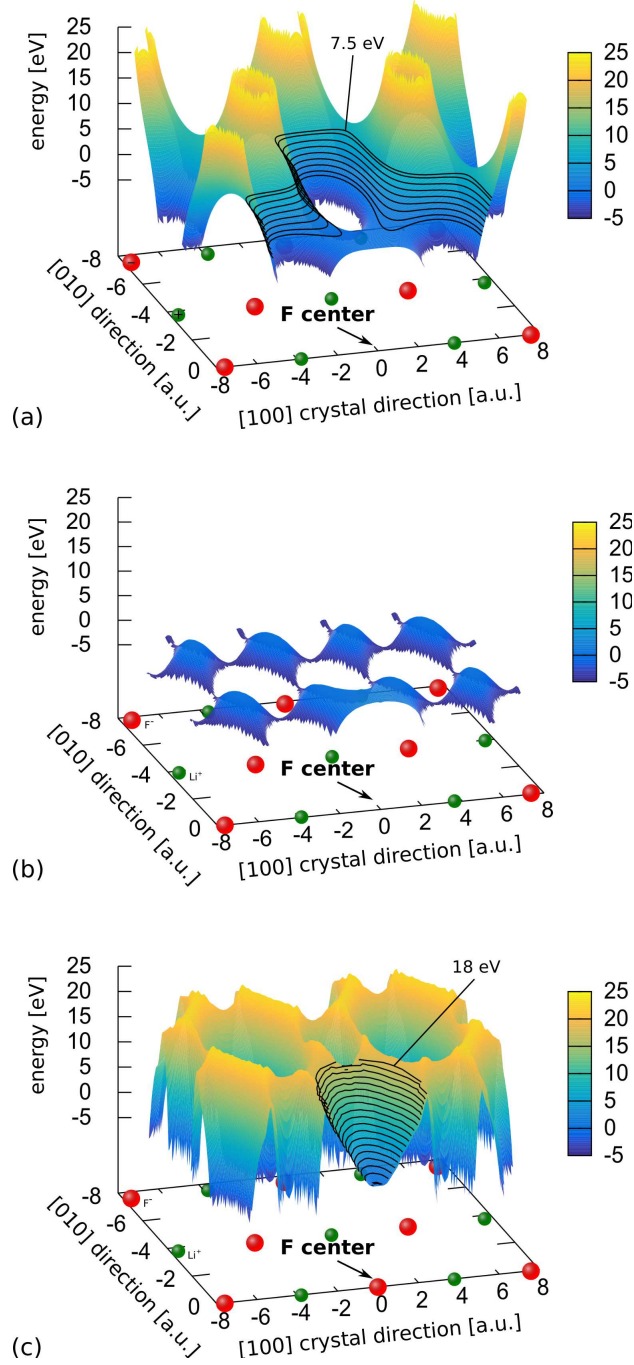}
\caption{(Color online) Two-dimensional cut through (a) the Madelung potential, (b) the effective Kohn--Sham potential $V_{KS}^{(LDA)}$ with the local density approximation to the exchange-correlation potential, and (c) the effective Kohn--Sham potential $V_{KS}^{(TB-mBJ)}$ using the TB-mBJ exchange-correlation potential at the F-center vacancy site, located at the origin, within the (001) crystal plane. The values of all potentials are set to zero at the origin, i.e., the vacancy site.}
\label{fig:potentials_3d}
\end{figure}

\begin{figure}[htb]
\centering \includegraphics[angle=0,width=0.5\columnwidth]{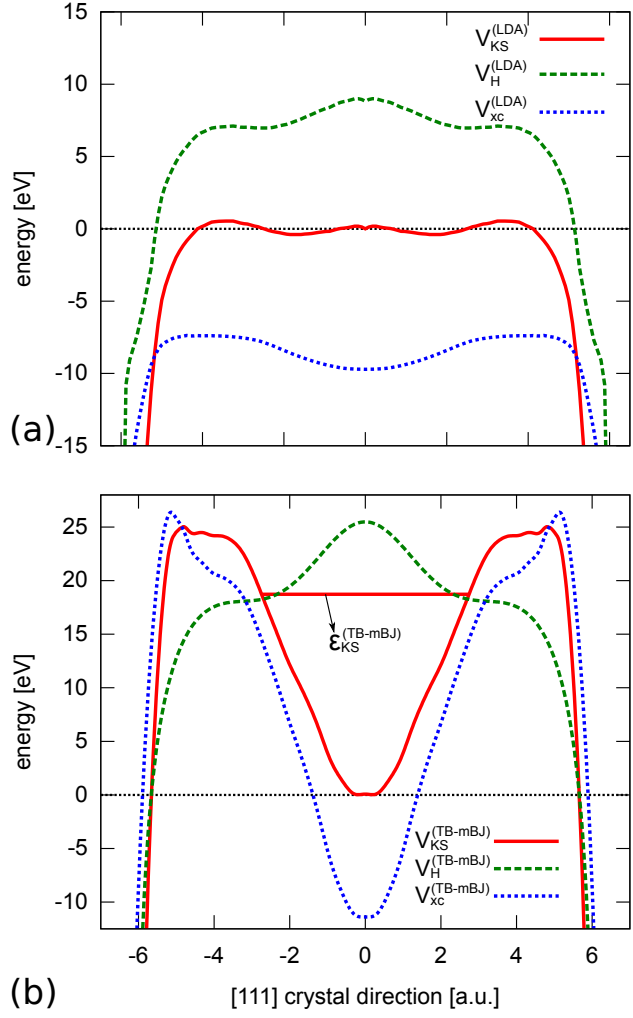}
\caption{(Color online) Cut through the effective Kohn--Sham potential V$_{KS}$ (red solid line) (a) in the local-density approximation and (b) using the TB-mBJ exchange-correlation potential in the F-center defect region of an LiF crystal along the [111] direction. Green (dashed) and blue (dotted) lines are the Hartree potential V$_{H}$ (including the Madelung potential) and the exchange-correlation potential V$_{xc}$, respectively. The horizontal line labeled with $\varepsilon_{KS}$ denotes the Kohn--Sham single-particle energy level of the localized F-center electron.}
\label{fig:pot_contributions}
\end{figure}

Shortly after the discovery of the Mollwo ($\sim a^{-2}$) and the Mollwo--Ivey ($\sim a^{-n}, n=1.81$) scaling for the F center absorption energies, two complementary simple models were proposed: Fr\"ohlich\cite{Fro33} deduced the $a^{-2}$ scaling from the energy spacing between bands in the model of a delocalized nearly free electron in a potential of period $2a$, $\Delta E = \hbar^2\pi^2/(2m^{*}a^2)$, where $m^{*}$ is the effective mass. St\"ockmann, instead, proposed a hard-wall box potential\cite{Sto51,Sto52-1,Sto52-2} for a completely localized defect electron that also gives rise to an $a^{-2}$ scaling. The hard wall potential was viewed as a simplification for the local potential originating from the electrostatic Madelung potential: due to the missing anion, there is a plateau-like almost constant potential in the space between the 6 neighboring cations and the repulsive potential of the surrounding anions forms the wall of the potential box that has an approximate diameter of the lattice constant $a$. A two-dimensional cut through the Madelung potential is shown in Fig. \ref{fig:potentials_3d} (a) where the black contour lines show the effective wall of about 8 eV around the four in-plane Li ions adjacent to the vacancy. This observation gives rise to the picture of the F-center defect electron as a particle in the box model frequently exploited and refined in subsequent works\cite{Fow68}. The two seemingly contradictory models of Fr\"ohlich and St\"ockmann yield the same exponent $n=2$ since both models treat, effectively, the kinetic energy of the electron, either confined to a periodic structure or to a hard-walled box. In turn, the systematic deviation to smaller values below 2, $n=1.8$, extracted by Ivey from a large data set was attributed first by Wood\cite{Woo65} within his ``point-ion model'' to the competition between the quantum-confined kinetic energy and the Madelung potential energy, $-\alpha_M/a$, where $\alpha_M$ is the crystal specific Madelung constant, which scales $\sim a^{-1}$. The Mollwo--Ivey relation was introduced as an effective single-exponent fit with $n=1.8$ to a power series expansion with the leading terms\cite{Woo65}
\begin{align}
\label{eq:Wood}
E_{abs} = c_{-2} a^{-2} + c_{-1} a^{-1}.
\end{align}
Small element-specific corrections to the point-ion model due to the influence of nearest neighbors such as orthogonalization effects, referred to as ion-size effects, were, in turn, considered to scale with higher inverse powers in $a$ [\onlinecite{Woo65}].
\\
An elegant alternative approach to Eq.\ \ref{eq:MI} was given by Malghani and Smith\cite{MalSmi92,SmiMal98} relating the absorption energy to the size of the defect wave function employing the Vinti sum rule\cite{Vin32}. The latter connects the moments of the frequency dependent absorption coefficient with the spatial extent of the wave function of the defect electron. The absorption energy can be related to the mean-squared radius $\langle s|r^2|s \rangle$ of the s-like ground state as
\begin{align}
\label{eq:energy_radius}
E_{abs} \approx \frac{3 \hbar^2}{2 m_e} \frac{1}{\langle s|r^2|s \rangle},
\end{align}
where $m_e$ is the mass of the electron. Eq.\ \ref{eq:energy_radius} becomes exact in the limit of the one-electron (independent particle) description and when the radiation field resonantly couples only two discrete states. To the extent that the radius of the wave function, $\langle s|r^2|s \rangle^{1/2}$, can be directly related to the lattice constant, Eq.\ \ref{eq:energy_radius} yields the Mollwo--Ivey relation. We will scrutinize the latter relationship in more detail below. Deviations from the exponent $n=2$ have been again attributed to the Madelung potential (see Eq.\ \ref{eq:Wood}). In turn, ion-size effects are invoked to account for small deviations from the Mollwo--Ivey exponent $n=1.8$.
\\
A critical analysis of these intuitive models proposed for the Mollwo--Ivey relation in the light of the ab-initio treatment of the problem leads to a revision of the particle-in-the-box picture. We have previously shown\cite{KarTiwLasTra14} for LiF that the F-center wave-function is localized within the vacancy site and has little weight at the neighboring ions. From the Madelung potential of Fig.\ \ref{fig:potentials_3d} (a) alone, one would expect a significant ``spill-out'' of the defect wave function to the nearest-neighbor cations. The strong confinement of the F-center electron is due to exchange interaction and the orthogonality of the wave function with respect to the ion-cores. In other words: the effective box size is smaller than suggested by the Madelung potential. These effects, sometimes called ``ion-size effects'', have been taken into account in the pre-ab-initio era through model potentials and trial wave-functions (see, e.g., Ref.\ [\onlinecite{Woo65}]).
\\
The Kohn--Sham potential (KS) $V_{KS}^{(LDA)}$ in the local-density approximation (Fig.\ref{fig:potentials_3d} (b)), obtained from a periodic DFT calculation, appears to lack any confining potential. This is because the KS potential is the common potential for all electrons. After accounting for core and the valence electrons, the defect electron, due to orthogonality constraints, is localized within the vacancy region. A cut through the ground-state wave-function of the F-center (see Fig.\ \ref{fig:pofI}) shows clearly a point of inflection which indicates the presence of an effective potential barrier and penetration into a ``classically forbidden'' region. Parenthetically we note that the PBE exchange-correlation functional\cite{PBE_ref} yields almost identical results for the total KS potential and the defect-electron wave function on the scale of Figs.\ \ref{fig:pofI} and \ref{fig:potentials_3d} (b).
\\
The potential barrier becomes visible (Fig.\ \ref{fig:potentials_3d} (c)) in the Kohn--Sham potential $V_{KS}^{(TB-mBJ)}$, calculated with the TB-mBJ approximation to the exchange-correlation potential. In this case, a pronounced potential well is present. Its depth varies between 17 and 25~eV depending on the crystallographic direction. A cut through the potential surface along the [111] direction (fig.\ \ref{fig:pot_contributions}) clearly shows that the Hartree potential, consisting of the electrostatic contributions of the point nuclei and the self-consistent electron density, gives rise to a local maximum in the potential near the location of the color center. Within the TB-mBJ approximation, it is the exchange-correlation potential that is responsible for the formation of an attractive potential well in the total effective Kohn--Sham potential and the confinement of the s-like defect-electron orbital with Kohn--Sham energy $\epsilon_{KS}^{(TB-mBJ)}(s)$ (Fig.\ \ref{fig:pot_contributions} (b)). The large differences between $V_{KS}^{(LDA)}$ and $V_{KS}^{(TB-mBJ)}$ are not surprising as the TB-mBJ exchange-correlation potential is designed to reproduce the experimental band gap of, e.g., LiF, which is severly underestimated by the LDA, by raising the effective potential in the interstitial regions. In view of these differences, however, an interpretation in terms of a realistic landscape should be taken with caution as the confinement of the wavefunction stems largely from the orthogonality requirement built into the non-local potential. We will use $V_{KS}^{(TB-mBJ)}$ in the following rather as a useful approximation to the effective potential that an isolated defect electron would feel in an otherwise fixed charge distribution.
\\
Despite the pronounced differences between the potentials $V_{KS}^{(LDA)}$ and $V_{KS}^{(TB-mBJ)}$ (Fig.\ \ref{fig:pot_contributions} (a) and (b), respectively), as well as the non-local Hartree--Fock potential the resulting wave-functions for the defect electron are very similar (Fig.\ \ref{fig:pofI}) featuring a point of inflection indicative for an effective confining potential well. Our current observations shed new light on Eq.\ \ref{eq:Wood}: what was earlier considered to be a small ``ion-size'' correction to the leading exponent $n=1.8$, is in fact the dominant contribution to the formation of the (effective) well and, thus, to the Mollwo--Ivey exponent itself. It should be noted, however, that the effective size of the potential well in terms of the classically allowed region of the wave function, as marked by the point of inflection, cannot be determined for all spatial directions. Along, e.g., [100] no point of inflection can be found in LDA and ROHF wavefunctions (Fig.\ \ref{fig:pofI} (b)). $\epsilon_{KS}^{(TB-mbJ)}$ lies above the barrier top in this direction, still the TB-mBJ wavefunction displays a point of inflection due to orthogonalization. We introduce in the following as effective measure for the size of the F-center function the position of the (first) zero of the s state along the [100] direction which accounts for the required orthogonality to the nearest-neighbor ionic states. Another measure of the size of the F-center is the mean-square radius of the wave-function $\langle s|r^2|s\rangle$ (Eq.~(\ref{eq:energy_radius})).
\\
The spill-out of the defect wave function into the region of neighboring ions (Fig.\ \ref{fig:pofI}) is experimentally accessible through the Fermi contact term $A_0$ in the hyperfine interaction between the defect-electron and the ion cores surrounding it. This quantity has been measured experimentally through electron-nuclear double resonance (ENDOR), developed by Feher\cite{Feh57} and it is proportional to the spin density $|\psi_{\uparrow}(0)|^2-|\psi_{\downarrow}(0)|^2$ at the nuclear sites of the host crystal\cite{Buc00}
\begin{align}
A_0 = -\frac{2}{3}\mu_0 g_e \beta_e g_n \beta_n (|\psi_{\uparrow}(0)|^2-\psi_{\downarrow}(0)|^2),
\end{align}
with the Bohr magneton $\beta_e$, the nuclear magneton $\beta_n$, the electron $g_e$ and nuclear $g_n$ factors, and the total density difference of spin up and spin down electrons $|\psi_{\uparrow}(0)|^2-|\psi_{\downarrow}(0)|^2$. Holton and Blum\cite{HolBlu62} performed systematic ENDOR studies and measured the Fermi contact term at the first, second, and third nearest neighbor ions of the F-center electron in various alkali halides. Mallia et al.\ [\onlinecite{MalOrlRoeUgl01}] as well as Leit\~ao et al.\ [\onlinecite{LeiCapVugBie02}] performed unrestricted Hartree--Fock (UHF) and spin-polarized LDA ab-initio calculations of $A_0$ which agree well with our data. UHF gives a Fermi contact term for the nearest-neighbor Li$^+$ ion in LiF to within one percent of the experimental value ($|\psi_{\uparrow}(0)|^2-\psi_{\downarrow}(0)|^2=0.0225$~bohr$^{-3}$) while $V_{KS}^{(LDA)}$ confines the defect electron too weakly resulting in an overestimation by $\sim 10$~\%. TB-mBJ, on the other hand, localizes the defect orbital too strongly leading to an underestimation of $A_0$ by $\sim 75$~\%. Independent of the method, the errors become larger for crystals with cations larger\cite{LeiCapVugBie02} than Li$^+$ as well as for second and third nearest neighbor sites\cite{MalOrlRoeUgl01} of the F center.
\\
Leit\~ao et al.\ found that an unrestricted or spin polarized calculation is a prerequisite for accurately determining the Fermi contact. In the ROHF approximation the spin density reduces to the density of the F-center electron and, therefore, it neglects spin-dependent polarization of core electrons. This polarization accounts for $\sim20$\% of the spin density at the nearest neighbor Li$^+$ site in LiF and up to $30$~\% in crystals\cite{LeiCapVugBie02} with larger ions. Despite this substantial difference between ROHF and UHF we find essentially identical single-particle defect wave functions for the F-center electron and identical Fermi contact terms if, in both approaches, only the defect orbital is considered. This indicates that $A_0$ is strongly influenced by local properties of the ions surrounding the defect and, therefore, is not a reliable measure for the confining potential. In the following we focus on ROHF defect wave functions as they are the starting point for the CASPT2(ROHF) calculations of the F-center absorption energies.

\section{Scaling with lattice constant}
\label{sec:scaling}
\begin{figure}[htb]
\centering \includegraphics[angle=0,width=0.5\columnwidth]{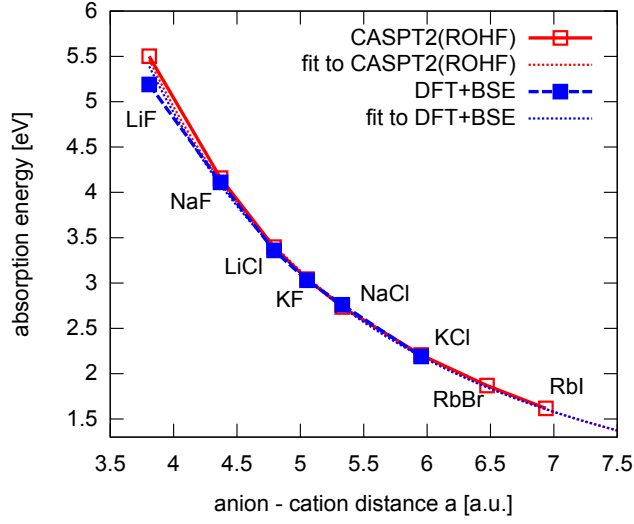}
\caption{(Color online) F-center absorption energies of the scaled alkali-halide model (scaled or ``stretched'' LiF) as a function of the anion-cation distance. The solid red line shows CASPT2 results, the dashed blue line are results from DFT+BSE calculations. Dotted red and blue lines are fits of Eq.\ \ref{eq:MI} to CASPT2 and DFT+BSE results, respectively. Effects of relaxation or electron-phonon coupling are not included. Fits yield Mollwo--Ivey exponents of $n_{CASPT2}=2.04$ and $n_{BSE}=2.01$. Due to the larger influence of ion-size effects on LiF(a$_{LiF}$) (smallest lattice constant) we neglect this point in the fits.}
\label{fig:energies_stretched}
\end{figure}

\begin{figure}[htb]
\centering \includegraphics[angle=0,width=0.5\columnwidth]{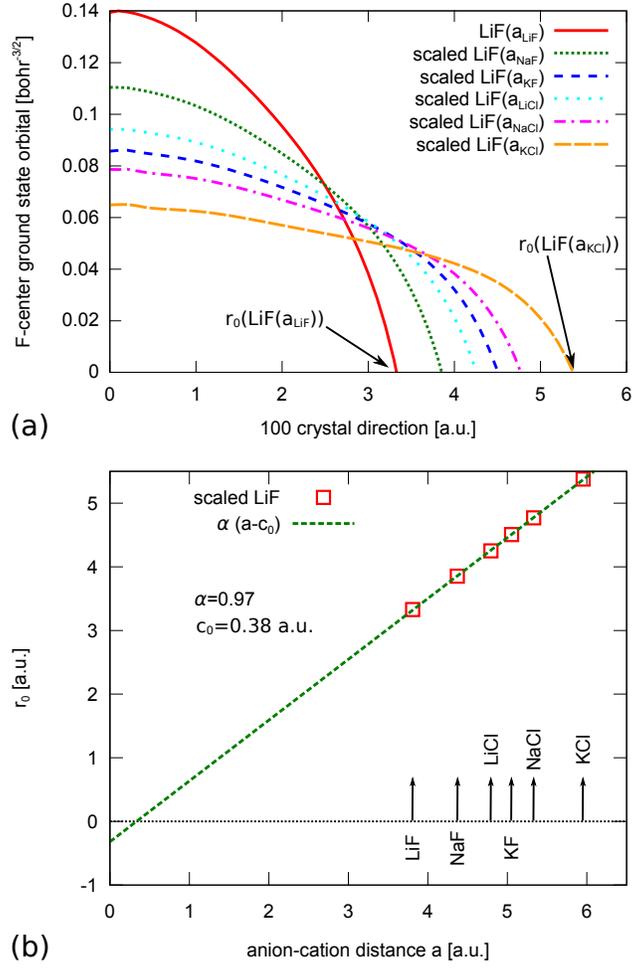}
\caption{(Color online) (a) Cut through the ROHF orbital $|s\rangle$ of the F-center electron in its ground state for the scaled LiF crystal evaluated at lattice constants $a$ corresponding to LiF, NaF, KF, LiCl, NaCl, and KCl. For clarity the wave functions are plotted only up to their first radial node $r_0$. (b) Variation of the position of the radial node $r_0$ with the scaled lattice parameter $a$.}
\label{fig:QC_orbs_stretched}
\end{figure}

To explore the physics underlying the Mollwo-Ivey relation, we introduce in the following the scaled alkali-halide model for which we determine its F-center absorption spectrum and wave functions fully ab-initio. In this model also referred to in the following as scaled LiF we treat the lattice parameter $a$ as continuous variable covering the lattice constants of all alkali halides while retaining the ionic constituents Li$^+$ and F$^-$ thereby disentangling the dependence on lattice constant from those on the ionic cores (i.e.\ ``ion-size effects''). We choose Li$^+$ and F$^-$ as the constituents as they represent the alkali halide with the smallest ionic radii coming closest to the ion-point model. Obviously, for the study of the scaling with $a$ presented in the following we omit any lattice relaxation or corrections due to electron-phonon coupling that result in corrections for real materials (see below). For scaled LiF, the F-center absorption energy (Fig.\ \ref{fig:energies_stretched}) decreases monotonically with $a$ resulting in an effective Mollwo--Ivey exponent remarkably close to $n=2$ (the CASPT2 calculation yields $n_{CASPT2}=2.04$, while the DFT+BSE approach yields $n_{BSE}=2.01$) almost perfectly matching the prediction for the particle in the box model. As measure for the extent of the defect wave function we use the position of the first radial node $r_0$ of $|s\rangle$ along the [100] crystal direction. As noted above, along the [100] direction the single-particle energy of $|s\rangle$ typically lies above the potential barrier which renders other measures such as, e.g., the point of inflection unreliable. For all lattice constants considered (Fig.\ \ref{fig:energies_stretched}) we find that $\sim 80$\% of the defect electron probability density is localized within a sphere with a radius of $r_0$ centered at the vacancy. Cuts through the $|s\rangle$ ROHF wave function along the [100] direction (Fig.\ \ref{fig:QC_orbs_stretched}) evaluated at anion-cation distances corresponding to those of various alkali halides (shown for clarity only for $r \leq r_0$) display a linear increase of $r_0$ (see Fig.\ \ref{fig:QC_orbs_stretched} (b)). The effective radius for the scaled system can be fitted to $r_0(a) = \alpha (a-c_0)$ with $\alpha=0.97$ and $c_0 = 0.38$~bohr representing the effective range of the exchange potential of the next-nearest cation. It is interesting to note that by applying the Vinti sum rule (Eq.\ \ref{eq:energy_radius}) to the experimental absorption spectra of real materials Malghani and Smith found a similar linear increase of the rms radius $\langle s|r^2|s \rangle^{1/2}$. From this observation they concluded that ion-size effects have only a negligible influence on the spatial extent of $|s\rangle$. This would suggest that also $r_0$ for real alkali halides would closely follow $r_0(a)$ of the scaled systems. We show in the following that this is not the case.

\section{Ion-size effects}
\label{sec:ion-size-effects}
\begin{figure}[htb]
\centering \includegraphics[angle=0,width=0.5\columnwidth]{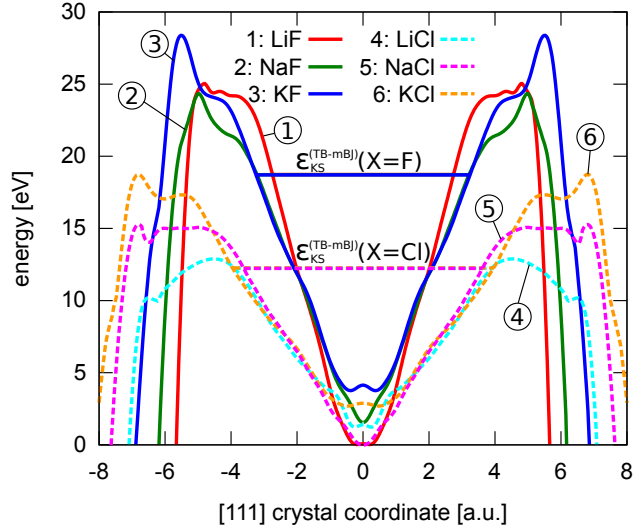}
\caption{(Color online) Cut through the effective Kohn--Sham potentials $V_{KS}^{(TB-mBJ)}$ in the F-center defect region in fluoride crystals (solid lines) and chlorides (dashed lines). Potentials are shifted in energy such that the single-particle Kohn--Sham energies $\epsilon_{KS}^{(TB-mBJ)}$ of the corresponding s-type defect orbitals coincide for fluorides ($\epsilon_{KS}^{(TB-mBJ)}(X=F)$ solid, upper horizontal line) and chlorides ($\epsilon_{KS}^{(TB-mBJ)}(X=Cl)$ dashed, lower horizontal line).}
\label{fig:DFT_pots}
\end{figure}

\begin{figure}[htb]
\centering \includegraphics[angle=0,width=0.5\columnwidth]{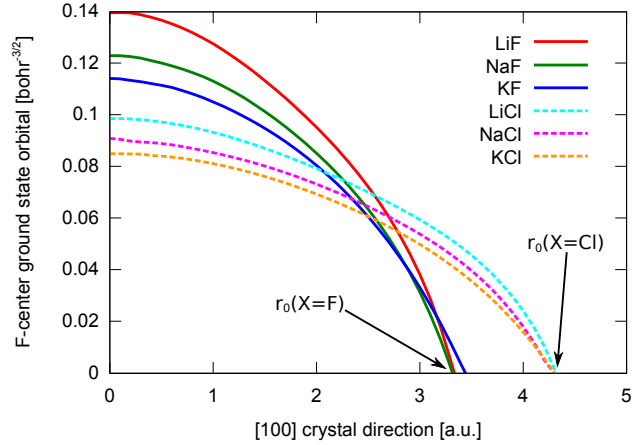}
\caption{(Color online) Cut through the ROHF orbital $|s\rangle$ of the F-center electron in its ground state for various alkali-halide crystals. For clarity the wave functions are plotted only up to their first radial node. Note the clustering of $r_0$ for a given halide $r_0(X)$.}
\label{fig:QC_orbs}
\end{figure}

\begin{figure}[htb]
\centering \includegraphics[angle=0,width=0.5\columnwidth]{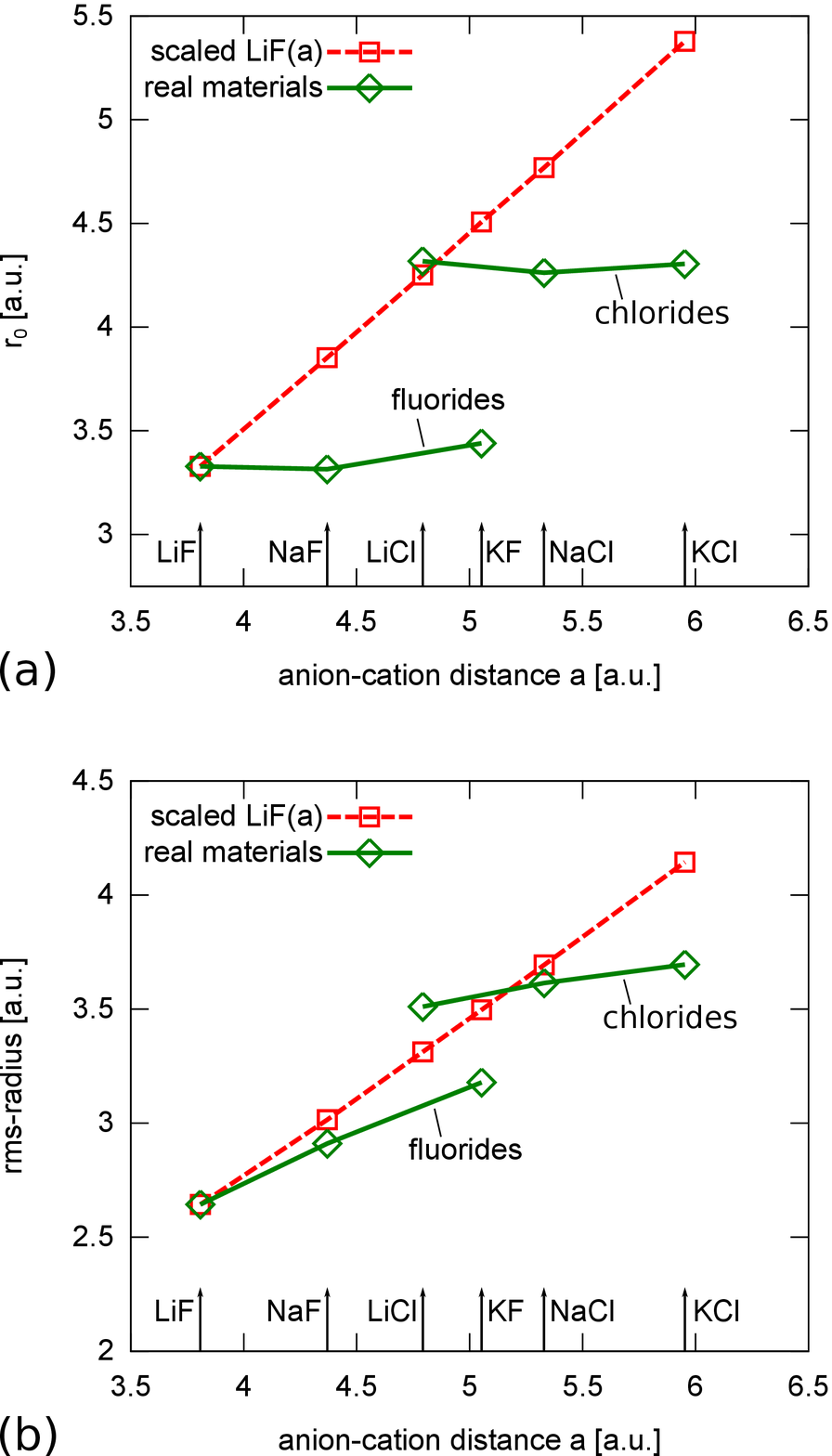}
\caption{(Color online) (a) Position of the first radial node of the ROHF orbital $|s\rangle$ of the F-center electron in its ground state along the [100] crystal direction plotted as a function of anion-cation distance in unrelaxed scaled LiF (dashed red line) and real unrelaxed alkali-halide crystals (solid green line). (b) Same as (a) but root-mean square (rms) radius $\langle s | r^2 | s \rangle^{1/2}$.}
\label{fig:r0_rms}
\end{figure}

\begin{figure}[htb]
\centering \includegraphics[angle=0,width=0.5\columnwidth]{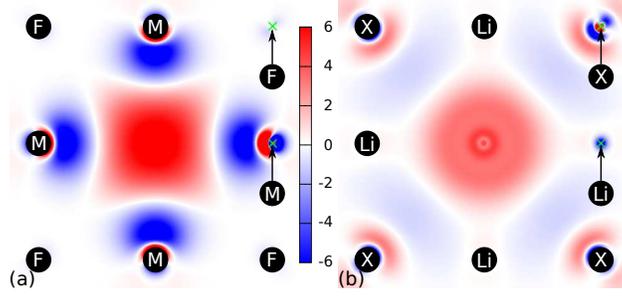}
\caption{(Color online) (a) Defect-electron density differences in $10^{-4}$/bohr$^3$ for the F-center between NaF and LiF ($\rho_{NaF}-\rho_{LiF}$) and (b) difference between LiCl and LiF ($\rho_{LiCl}-\rho_{LiF}$) for a constant anion-cation distance of 5.95~bohr. Replacing Li and F ions with the larger Na and Cl ions, respectively, compresses the defect-electron density in the vacancy.}
\label{fig:delta_HForbs}
\end{figure}

We turn now to the real alkali halides. The effective Kohn--Sham potential wells (Fig.\ \ref{fig:DFT_pots}, shifted in energy for clarity) form distinct groups depending on the anionic species. Within the group of fluoride and chloride crystals, respectively, all potential wells, independent of the lattice parameter, have a similar depth and similar slope of the the walls. We find the same trend for the spatial extent of the F-center ground state ROHF orbitals for the different, unrelaxed alkali halides (Fig.\ \ref{fig:QC_orbs}). Typically, 75 to 90\% of the electron are localized within a sphere around the defect with a radius $r_0$. The defect orbitals clearly split into two groups according to the anion species. Remarkably, the radial nodes for a given anion species almost coincide. We find for fluorides a radial node at $r^F_0 \approx 3.4$~bohr and for chlorides $r^{Cl}_0 \approx 4.3$~bohr. We have previously shown\cite{KarTiwLasTra14} that for LiF the displacement of the Li cations due to relaxation is  $\approx 0.1$~bohr. We, therefore, expect that ordering into groups with almost identical $r_0$ depending on the anion, $r_{0,X}$, persists when material-specific relaxation is included. Likewise, the close correspondence between the HF defect orbitals and the natural orbital of the N-electron wave function evaluated at the CASPT2 level, demonstrated for LiF [\onlinecite{KarTiwLasTra14}], indicates that correlation effects will not lift this degeneracy of $r_{0,X}$ with respect to the cationic constituent M.
\\
The dramatic difference between the linear increase in $r_0$ for the scaled LiF model, $r_0(a)$, and the discontinuous jumps in real alkali halides is demonstrated in Fig.\ \ref{fig:r0_rms} (a). This clustering is not specific to $r_0$. A similar, yet less pronounced, clustering and deviation from linear scaling is found for the rms radius $\langle s|r^2|s \rangle^{1/2}$ (Fig.\ \ref{fig:r0_rms} (b)). Both $r_0$ and rms display strong ion-size effects suggesting that larger ions tend to compress the ground-state wave function in line with results found by Gourary and Adrian\cite{GouAdr56} who added effects of exchange to their point-ion model. This differs from the linear scaling and negligible influence of ion-size effects previously suggested\cite{MalSmi92,SmiMal98}.
\\
The element-specific effects of the ionic core on the defect state can be studied on the ab-initio level by considering an alkali-halide model system with \textit{fixed} lattice constant $a$ while varying the cation and anion species. We choose as fixed anion-cation distance $a_{KCl}/2= 5.95$~bohr. In the following $M \in \{Li,Na,K\}$ denotes the alkali metal, $X \in \{F,Cl\}$ the halide and $\rho_{MX}$ denotes the F-center electron density in the crystal MX. We study the density variation within the fixed F-center volume. The effect of exchanging the metal, $\rho_{NaF}-\rho_{LiF}$, and the halide, $\rho_{LiCl}-\rho_{LiF}$, are twofold (Fig.\ \ref{fig:delta_HForbs}). First, in the vicinity of the missing anion $\rho_{NaF}-\rho_{LiF}$ and $\rho_{LiCl}-\rho_{LiF}$ show an accumulation (or compression) of the defect-electron density within the vacancy which displays strong directionality. Replacing the Li cation by an Na cation (Fig.\ \ref{fig:delta_HForbs} (a)) strongly compresses the defect electron in the vacancy along the [100] and the [010] crystal direction. The direction of the compression is along the [110] and [1-10] crystal axis, the two diagonals in Fig.\ \ref{fig:delta_HForbs} (b), when the F anion is replaced by a Cl anion and is weaker than for the exchange of cations. Second, at the position of the ionic neighbor to be exchanged we also observe an increase in density when smaller ionic cores are replaced by larger ones. The ROHF Fermi contact term at the corresponding metal or halide site increases by a factor $\sim3$ upon replacing Li$^+$ by Na$^+$ or F$^-$ by Cl$^-$. This increase of density reflects the increasingly stringent orthogonality requirement on the defect wave function imposed by large ionic cores. Thus, the density increases both at the vacancy site and at the neighboring ionic site. While the former is the consequence of the compression of the defect wave function, the latter is a subtle local effect on the tails of the wave function due orthogonalization and should not be taken as a measure for the effective size of the confining ``box'' potential. The compression observed in the present study is in contradiction to the results of Smith and Inokuti\cite{SmiIno01} who concluded that the rms of the F-center wave function increases with the ionic size of the cation. For the model alkali halide system with fixed lattice constant $a_{KCl}$ we find a decrease of the rms value from 4.14 to 4.04~bohr when replacing LiF by NaF and from 4.18 to 4.04~bohr when replacing LiCl by NaCl. We note that the rms radii are almost independent of the anionic constituent.
\\
The approximate independence of $r_0$ (or rms) on the anionic species in real alkali halides (Figs.\ \ref{fig:r0_rms}) can be explained in terms of the well-known model of effective ionic radii in pristine crystals\cite{Sha76}. From measurements of the lattice constants in different alkali-halide crystals radii for anions and cations are extracted which are expected to be independent of the crystal composition. Adding up the effective radii $r_M$ and $r_X$ of the cation M and the anion X gives, to a good degree of approximation, the anion-cation distance $a_{MX}$ of the crystal MX. Following this reasoning, the F-center electron should have the same spatial extent within a given halide which is in qualitative agreement with the $r_0$ values of the F-center in real materials (Fig.\ \ref{fig:r0_rms}). Using this effective ionic radii model, we find even quantitative agreement when we estimate the $r_0$ value of the real material from these of the scaled LiF value $r_0(a_{MX})$ after correcting for the ionic radius of the cation M. The second column in table \ref{tab:zero-crossings} lists the $r_0(a_{MX})$ value evaluated within the scaled LiF model at the lattice constant $a_{MX}$. The $r_0$ value of the real alkali halide MX, $r_{0,MX}$, follows then as
\begin{align}
\label{eq:r0_corrected}
r_{0,MX} = r_0(a_{MX}) - (r_M - r_{Li})
\end{align}
by correcting for the difference between the ionic radii for the metal under consideration (third column) and for Li chosen as the metallic constituent of the scaled alkali halide. The resulting values (Eq.\ \ref{eq:r0_corrected}, fourth column) provide a near-perfect match with the numerical values $r_0$ found in our ab-initio simulation (Fig.\ \ref{fig:r0_rms}). This accurate prediction of the effective-radii model for $r_0$ suggests the following conclusions: along the [100] direction the extent of the ground state defect wave function in real materials is constant for a given anionic species because the increase of the vacancy wave function due to growing lattice constant when changing M, e.g., from LiF to NaF is completely compensated for by the increased size of the cation, i.e.\ the nearest neighbor size. When exchanging the anion for given cation, going from, e.g., LiF to LiCl, the defect-wave function increases along the [100] direction exactly by the difference in anion radii, $r_{Cl^-}-r_{F^-}$. This estimate for $r_0$ perfectly matches the difference in anion-cation distance of $0.9$~bohr between crystals with the same cation but different anions F$^-$ and Cl$^-$.
\begin{table}
\begin{tabular}{c|c|c|c}
\hline
Crystal MX  & $r_0(a_{MX})$  & $r_M$  & $r_0(a_{MX}) - (r_M-r_{Li})$ \\
\hline
LiF         &$3.33$            &$1.70$        &$3.33$  \\
NaF         &$3.85$            &$2.19$        &$3.36$  \\
KF          &$4.51$            &$2.87$        &$3.34$  \\
LiCl        &$4.25$            &$1.70$        &$4.25$  \\
NaCl        &$4.77$            &$2.19$        &$4.28$  \\
KCl         &$5.38$            &$2.87$        &$4.21$  \\
\hline
\end{tabular}
\caption{\label{tab:zero-crossings} Correction of the position of the radial node $r_0$ of the defect wave function of scaled LiF due to the effective size of the real cation neighbor M. The second column lists the position of the first radial node along the [100] crystal direction of the ground-state defect wave function in scaled LiF, $r_0(a_{MX})$ with the lattice constant of the material MX. The third column gives the effective cation radii\cite{Sha76} $r_M$ and the forth column lists the corrected position of the radial node $r_{0,MX} = r_0(a_{MX}) - (r_M-r_{Li})$. All distances are given in bohr.}
\end{table}

\section{Absorption Energies}
\label{sec:energies}
\begin{figure}[htb]
\centering \includegraphics[angle=0,width=0.45\columnwidth]{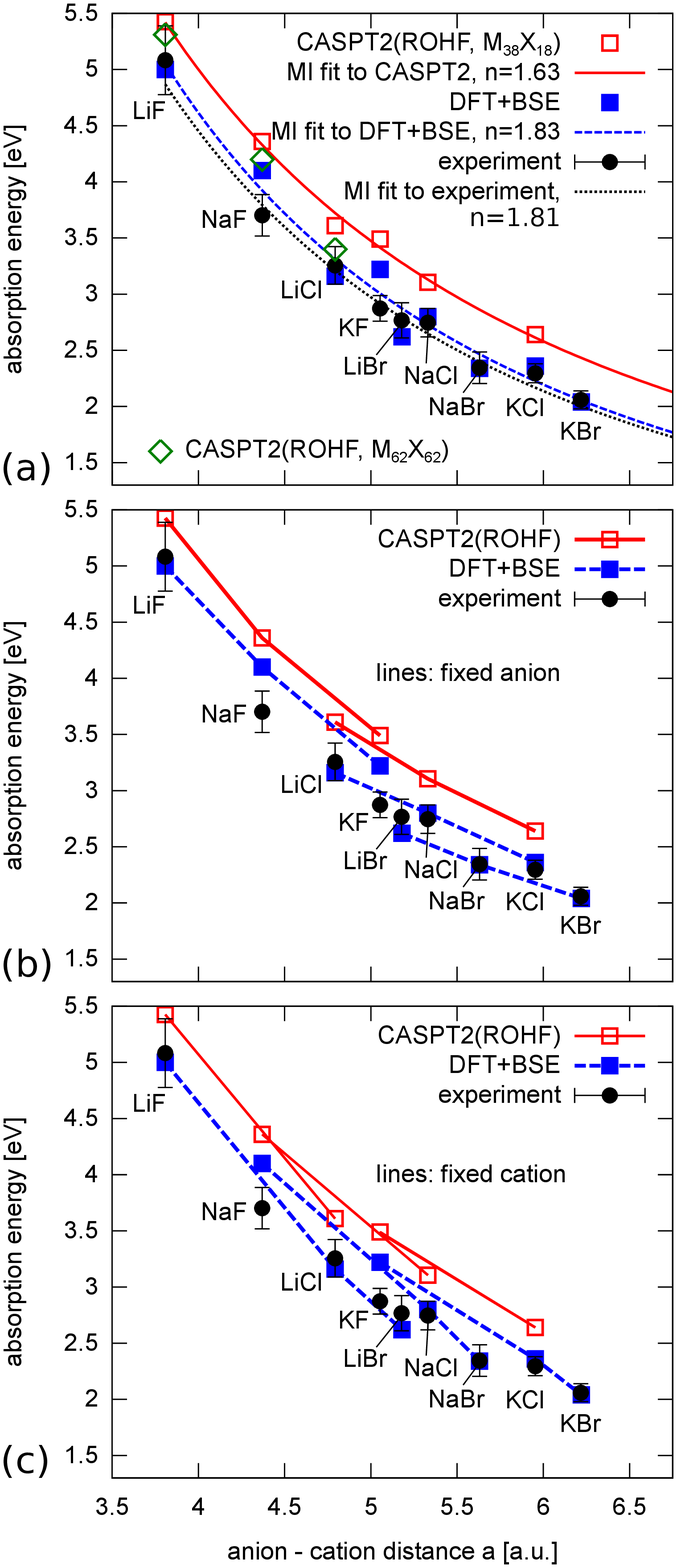}
\caption{(Color online) (a) F-center absorption energies of various alkali-halide crystals as a function of the anion-cation distance. Open-red squares are CASPT2(ROHF) results obtained within the embedded cluster approach (active cluster: M$_{38}$X$_{18}$), open-green diamonds are CASPT2(ROHF) results for an M$_{62}$X$_{62}$ embedded cluster, full-blue squares are DFT+BSE results obtained within a periodic boundary calculation. Black dots are experimental data\cite{DawPoo69}, the error bars indicate the full width at half maximum of the F-center absorption peak. In panel (a) Mollwo--Ivey-type fits (MI fits) to the theoretical and experimental data are plotted with Mollwo--Ivey exponents of $n_{CASPT2}=1.63$, $n_{BSE}=1.83$, and $n_{exp}=1.81$. In panels (b) and (c) theoretical absorption energies of crystals containing the same anionic and cationic species, respectively, are connected.}
\label{fig:energies_real}
\end{figure}

Turning now to the absorption energies of F centers in real alkali halides, their values will be determined by two competing tendencies, the lattice spacing and ion size. Since the absorption energies are universally proportional to the mean-square radius of the F-center ground-state wave function (Eq.\ \ref{eq:energy_radius}) we expect for $E_{abs}$ of the alkali halide MX an exponent systematically below two, $n < 2$, as the increased ion-size effects offset the increased lattice constant. Therefore, the reduction from $n=2$ to the fitted value $n=1.8$ itself is the signature of ion-size effects and not the subtle deviations from this exponent as previously thought. The Madelung potential, previously involved in the explanation of the deviation of the exponent from 2 (Eq.\ \ref{eq:Wood}), is of no relevance for the deviation of the Mollwo-Ivey exponent from 2. In the absence of ion-size effects but in the presence of the Madelung potential, the exponent would be $n=2$ as the ab-initio results for the scaled LiF model unambiguously show.
\\
Fig.\ \ref{fig:energies_real} (a) shows the experimental absorption energies\cite{DawPoo69} of the F-center in several materials together with the energies obtained from embedded quantum-chemistry cluster calculations and from DFT+BSE calculations. The error bars on the experimental data points indicate the large full-width at half maximum of the F-center absorption peak. Lines through the data points are Mollwo--Ivey-type fits with Mollwo--Ivey exponents of $n_{CASPT2}=1.63$ and $n_{BSE}=1.83$. These values lie well below 2 as expected from the analysis above and $n_{BSE}$ is quite close to experiment, $n_{exp}=1.81 \pm 0.10$. All theoretical data points (CASPT2 and DFT+BSE) are calculated for a relaxed geometry which is determined by a periodic boundary DFT calculation using the PBE exchange-correlation potential. Corrections due to electron-phonon interaction are not included as downward shifts of the absorption energies of below 0.1~eV, close to those in LiF \onlinecite{KarTiwLasTra14}], are too small to alter the scaling behavior of the absorption energies. The CASPT2 absorption energies, denoted by the open-red squares (Fig.\ \ref{fig:energies_real}), are determined within an M$_{38}$X$_{18}$ cluster and are extrapolated to the converged basis set limit (see [\onlinecite{KarTiwLasTra14}]). For LiF, NaF, and LiF calculations with an active M$_{62}$X$_{62}$ cluster are computationally feasible leading to downward shifts of $E_{abs}$ between 0.12~eV (LiF) and 0.21~eV (LiCl). M$_{62}$X$_{62}$ absorption energies are denoted by open-green diamonds (Fig.\ \ref{fig:energies_real} (a)).
\\
In order to illustrate that the effective Mollwo--Ivey line represents a ``smoothed'' average over discrete ion-size effects we present in Fig.\ \ref{fig:energies_real} (b) and (c) the same data, however connected by lines according to the same anionic (Fig.\ \ref{fig:energies_real} (b)) and cationic (Fig.\ \ref{fig:energies_real} (c)) constituent. Ion-specific offsets and variations of the slope resembling the $r_0$ and rms values (Fig.\ \ref{fig:r0_rms}) are clearly observable even though more pronounced in the simulations than in the experiment. Fitting the data separately for every anion with a Mollwo--Ivey type relation leads to exponents smaller than the one obtained by a fit of all absorption energies. When the cation species is kept constant a steeper decrease is found because the size of the narrowest constriction of the vacancy due to the nearest-neighbor cations continuously increases with $a$. We note that in an earlier work by Smakula\cite{SmaMayRep62} (see also [\onlinecite{Woo65}]) separate fits to the Mollwo--Ivey relation for alkali halides with the cation fixed were presented. However, because of the weak dependence of the experimental data final conclusions were difficult to draw.

\section{Conclusions}
\label{sec:conclusions}
We present in this work an ab-initio study of the Mollwo--Ivey relation of F-centers in alkali-halide crystals based on post-DFT and post-HF methods. In contrast to earlier interpretations which stress the importance of the Madelung potential, we find ion-size effects to be the predominant mechanism forming the effective potential well within which the defect electron is mostly localized. The sizes of the neighbor ions determine the shape of the defect-electron wave function and are, therefore, responsible for the fractional Mollwo--Ivey exponent of 1.81. If it were not for ion-size effects a Mollwo--Ivey exponent of $n=2$ would emerge. We have introduced the model system of scaled LiF in which we increase the anion-cation distance while keeping ion sizes constant. The F-center absorption energies in scaled LiF obey a Mollwo--Ivey relation with an exponent of 2 equal to the three-dimensional particle-in-a-box model by St\"ockmann. The reduced Mollwo--Ivey exponent for ``real'' materials suggests neighboring ions compressing the defect electron wave function within the vacancy region. This leads to a reduced growth of the wave function's extent with increasing lattice parameter which, via the Vinti sum rule, is directly connected to the absorption energy. A qualitative picture of this compression is gained by studying ab-initio differences of defect-electron densities in different alkali-halides at a fixed anion-cation distance. A semi-quantitative picture is obtained by examining radial nodes of Hartree--Fock defect electron wave functions the positions of which perfectly agree with predictions from the effective ion-radii model.
\\
We have found ion-dependent offsets in the F-center absorption energies in our simulations also visible in experiment, however less pronounced. Future work should focus on effects responsible for the ``smoothing out'' of these offsets such as lattice relaxations, ion-dependent electron-phonon coupling corrections and other possible effects. Further, we hope the analysis presented above can be applied to color-centers in more complex materials and to defects hosting more than a single electron. Mollwo--Ivey like relations exist for such defects (see, e.g., [\onlinecite{WilMarWilKab77,LIWilAbe13}]), however, with exponents far from $n \approx 2$. The same is true for fluorescence from F centers where the Mollwo--Ivey like relations differ strongly\cite{Bal91,DaS05} from the absorption case due to the large Stokes shifts.

\section{Acknowledgments}
We thank H.\ Jiang and F.\ Aquilante for discussions and technical support. Further, thanks are due to L.\ Seijo for providing AIMPs for LiCl. This work was supported by the Austrian Fonds zur F\"orderung der wissenschaftlichen Forschung (Projects SFB F41 ``ViCoM'' and the doctoral college W1243 ``Solids4Fun''). F.K.\ and P.B.\ acknowledge the support by the TU Vienna doctoral college COMPMAT. L.W.\ acknowledges support by the National Research Fund, Luxembourg (Project C12/MS/3987081/TSDSN). Calculations were performed on the Vienna Scientific Cluster (VSC), on the SGI Altix UltraViolet 1000 of the Austrian Center for Scientific Computing (ACSC) installed at the Johannes Kepler University Linz and at the
IDRIS supercomputing center, Orsay (Proj. No. 091827).

\bibliography{library}

\newpage

\newpage

\newpage

\newpage

\newpage

\newpage

\newpage

\newpage

\newpage

\newpage

\end{document}